\documentclass[twocolumn, prl]{revtex4-1}

\usepackage{graphicx}
\usepackage{dcolumn}
\usepackage{siunitx}
\usepackage{bm}
\usepackage{amsmath}
\usepackage{amssymb}
\usepackage{mhchem}
\usepackage{hyperref}

\usepackage{xr}

\newcommand{\me}{\mathrm{e}}

\newcommand{\concat}{\oplus}

\begin{document}


\title{Crystal Graph Convolutional Neural Networks for an Accurate and\\
Interpretable Prediction of Material Properties}


\author{Tian Xie}
\author{Jeffrey C. Grossman}
\affiliation{Department of Materials Science and Engineering, Massachusetts Institute of Technology, Cambridge, Massachusetts 02139, United States}


\date{\today}

\begin{abstract}
The use of machine learning methods for accelerating the design of crystalline materials usually requires manually constructed feature vectors or complex transformation of atom coordinates to input the crystal structure, which either constrains the model to certain crystal types or makes it difficult to provide chemical insights. Here, we develop a crystal graph convolutional neural networks (CGCNN) framework to directly learn material properties from the connection of atoms in the crystal, providing a universal and interpretable representation of crystalline materials. Our method provides a highly accurate prediction of DFT calculated properties for 8 different properties of crystals with various structure types and compositions after trained with $10^4$ data points. Further, our framework is interpretable because one can extract the contributions from local chemical environments to global properties. Using an example of perovskites, we show how this information can be utilized to discover empirical rules for materials design. 
\end{abstract}


\maketitle



Machine learning (ML) methods are becoming increasingly popular in accelerating the design of new materials by predicting material properties with accuracy close to \textit{ab-initio} calculations, but with computational speeds orders of magnitude faster\cite{seko2015prediction, faber2016machine, xue2016accelerated}. The arbitrary size of crystal systems poses a challenge as they need to be represented as a fixed length vector in order to be compatible with most ML algorithms. This problem is usually resolved by manually constructing fixed-length feature vectors using simple material properties\cite{isayev2017universal, seko2015prediction, xue2016accelerated, ghiringhelli2015big, isayev2015materials} or designing symmetry-invariant transformations of atom coordinates\cite{schutt2014represent, faber2015crystal, seko2017representation}. However, the former requires case-by-case design for predicting different properties and the latter makes it hard to interpret the models as a result of the complex transformations.

In this letter, we present a generalized crystal graph convolutional neural networks (CGCNN) framework for representing periodic crystal systems that provides both material property prediction with DFT accuracy and atomic level chemical insights. Recent advances in ``deep learning" have enabled learning from a very raw representation of data, e.g.\ pixels of an image, making it possible to build general models that outperforms traditionally expert designed representations\cite{lecun2015deep}. By looking into the simplest form of crystal representation, i.e.\ the connection of atoms in the crystal, we directly build convolutional neural networks on top of crystal graphs generated from crystal structures. The CGCNN achieves similar accuracy with respect to DFT calculations as DFT compared with experimental data for eight different properties after being trained with data from the Materials Project\cite{jain2013commentary}, indicating the generality of this method. We also demonstrate the interpretability of CGCNN by extracting the energy of each site in the perovskite structure from the total energy, an example of learning the contribution of local chemical environments to the global property. The empirical rules generalized from the results are consistent with the common knowledge for discovering more stable perovskites and can significantly reduce the search space for high throughput screening.

\begin{figure}[b]
  \centering
  \includegraphics[width=\linewidth]{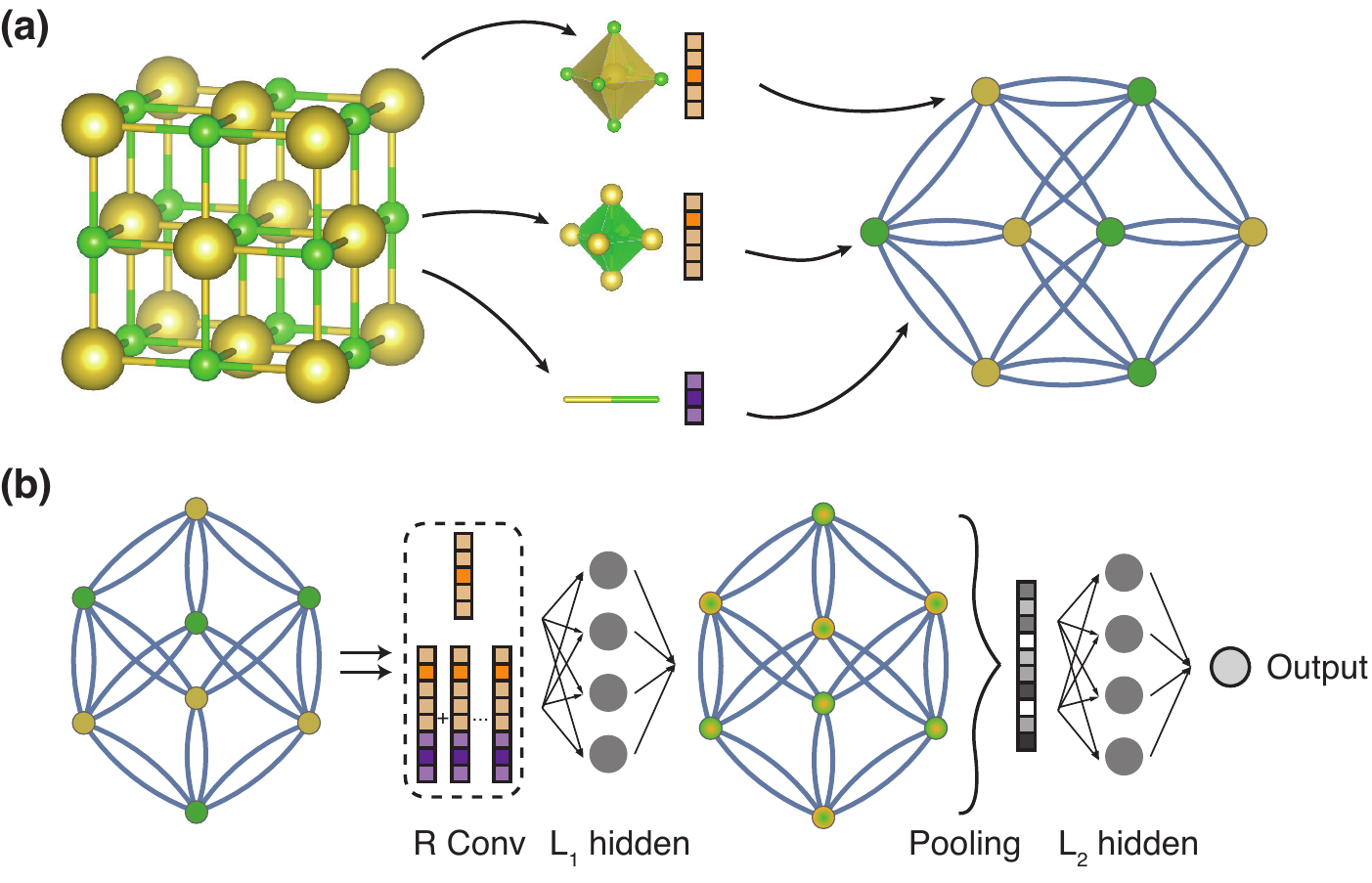}
  \caption{Illustration of the crystal graph convolutional neural network (CGCNN). (a) Construction of the crystal graph. Crystals are converted to graphs with nodes representing atoms in the unit cell and edges representing atom connections. Nodes and edges are characterized by vectors corresponding to the atoms and bonds in the crystal, respectively. (b) Structure of the convolutional neural network on top of the crystal graph. $R$ convolutional layers and $L_1$ hidden layers are built on top of each node, resulting in a new graph with each node representing the local environment of each atom. After pooling, a vector representing the entire crystal is connected to $L_2$ hidden layers, followed by the output layer to provide the prediction. }
  \label{fig:illustration}
\end{figure}

The main idea in our approach is to represent the crystal structure by a crystal graph that encodes both atomic information and bonding interactions between atoms, and then build a convolutional neural network on top of the graph to automatically extract representations that are optimum for predicting target properties by training with DFT calculated data. As illustrated in Figure \ref{fig:illustration} (a), a crystal graph $\mathcal{G}$ is an undirected multigraph which is defined by nodes representing atoms and edges representing connections between atoms in a crystal (the method for determining atom connectivity is explained in Supplemental Material\footnote{See Supplemental Material for further details, including Refs.~\cite{sanderson1951interpretation, sanderson1952explanation, cordero2008covalent, kramida2013nist, haynes2014crc, kingma2014adam, srivastava2014dropout, de2015charting, isayev2017universal, blatov2004voronoi}}). The crystal graph is unlike normal graphs since it allows multiple edges between the same pair of end nodes, a characteristic for crystal graphs due to their periodicity, in contrast to molecular graphs. Each node $i$ is represented by a feature vector $\bm{v}_i$, encoding the property of the atom corresponding to node $i$. Similarly, each edge $(i, j)_k$ is represented by a feature vector $\bm{u}_{(i, j)_k}$ corresponding to the $k$-th bond connecting atom $i$ and atom $j$.

The convolutional neural networks built on top of the crystal graph consists of two major components: convolutional layers and pooling layers. Similar architectures have been used for computer vision\cite{krizhevsky2012imagenet}, natural language processing\cite{collobert2008unified}, molecular fingerprinting\cite{duvenaud2015convolutional}, and general graph-structured data\cite{henaff2015deep, GilmerSRVD17} but not for crystal property prediction to the best of our knowledge. The convolutional layers iteratively update the atom feature vector $\bm{v}_i$ by ``convolution'' with surrounding atoms and bonds with a non-linear graph convolution function.
\begin{equation} \label{eq:conv}
	\bm{v}_i^{(t+1)} = \textrm{Conv}\left( \bm{v}_i^{(t)}, \bm{v}_j^{(t)}, \bm{u}_{(i, j)_k} \right), (i, j)_k \in \mathcal{G}
\end{equation}
After $R$ convolutions, the network automatically learns the feature vector $\bm{v}_i^{(R)}$ for each atom by iteratively including its surrounding environment. The pooling layer is then used for producing an overall feature vector $\bm{v}_c$ for the crystal, which can be represented by a pooling function,
\begin{equation}
  \bm{v_c} = \textrm{Pool} \left(\bm{v}_0^{(0)}, \bm{v}_1^{(0)}, ..., \bm{v}_N^{(0)}, ..., \bm{v}_N^{(R)} \right)
\end{equation}
that satisfies permutational invariance with respect to atom indexing and size invariance with respect to unit cell choice. In this work, a normalized summation is used as the pooling function for simplicity but other functions can also be used. In addition to the convolutional and pooling layers, two fully-connected hidden layers with the depth of $L_1$ and $L_2$ are added to capture the complex mapping between crystal structure and property. Finally, an output layer is used to connect the $L_2$ hidden layer to predict the target property $\hat{y}$. 

The training is performed by minimizing the difference between the predicted property $\hat{y}$ and the DFT calculated property $y$, defined by a cost function $J(y, \hat{y})$. The whole CGCNN can be considered as a function $f$ parameterized by weights $\bm{W}$ that maps a crystal $\mathcal{C}$ to the target property $\hat{y}$. Using backpropagation and stochastic gradient descent (SGD), we can solve the following optimization problem by iteratively updating the weights with DFT calculated data,
\begin{equation}\label{eq:optimization}
	\displaystyle{\min_{\bm{W}} J(y, f(\mathcal{C}; \bm{W}))}
\end{equation}
the learned weights can then be used to predict material properties and provide chemical insights for future materials design. 

In Supplemental Material (SM), we use a simple example to illustrate how a CGCNN composed of one linear convolution layer and one pooling layer can differentiate two crystal structures. With multiple convolution layers, pooling layers and hidden layers, CGCNN can extract any structure differences based on the atom connections and discover the underlaying relations between structure and property.

To demonstrate the generality of the CGCNN, we train the model using calculated properties from the Materials Project\cite{jain2013commentary}. We focus on two types of generality in this work: (1) The structure types and chemical compositions for which our model can be applied, and (2) the number of properties that our model can accurately predict.

\begin{figure}[tb]
  \centering
  \includegraphics[width=\linewidth]{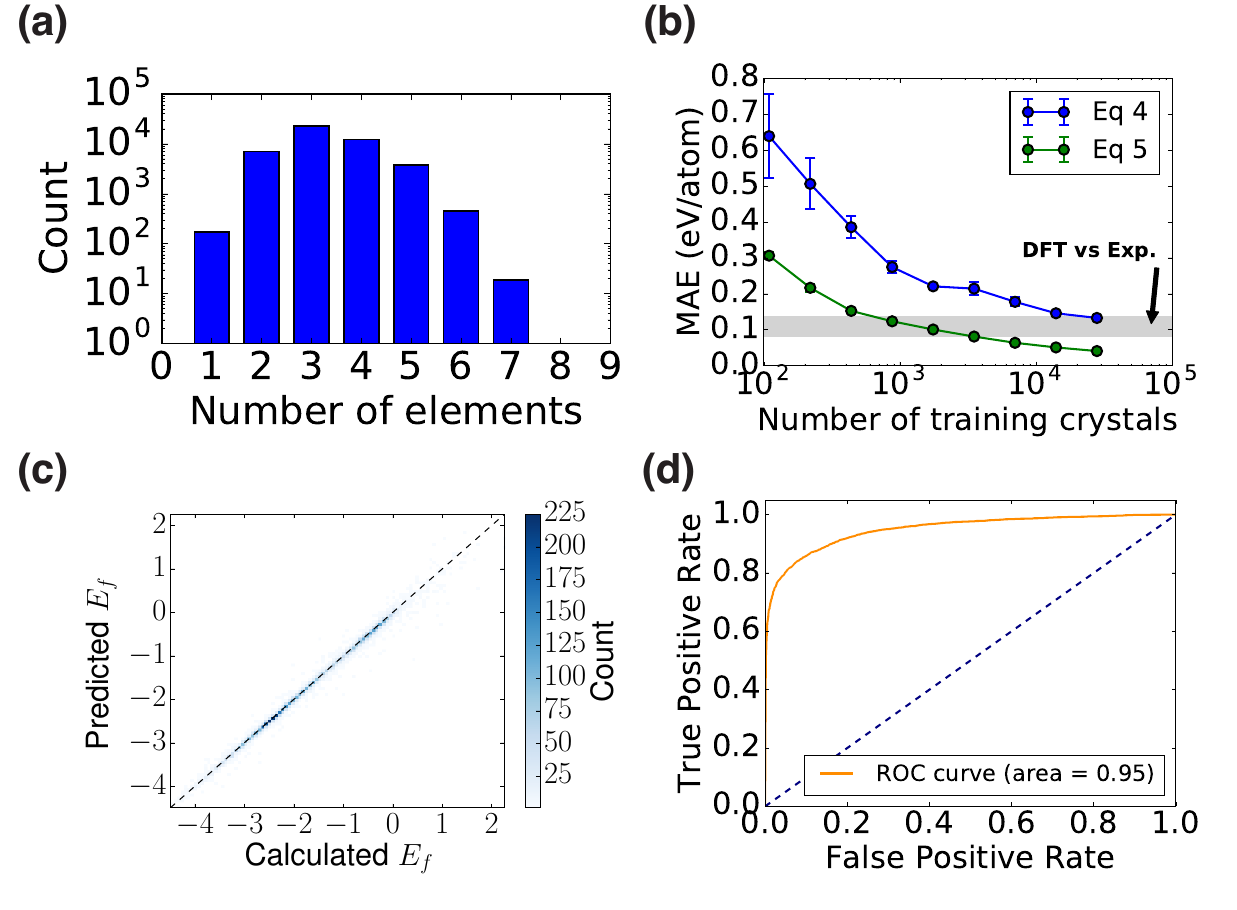}
  \caption{The performance of CGCNN on the Materials Project database\cite{jain2013commentary}. (a) Histogram representing the distribution of the number of elements in each crystal. (b) Mean absolute error (MAE) as a function of training crystals for predicting formation energy per atom using different convolution functions. The shaded area denotes the MAE of DFT calculation compared with experiments\cite{kirklin2015open}. (c) 2D histogram representing the predicted formation per atom against DFT calculated value. (d) Receiver operating characteristic (ROC) curve visualizing the result of metal-semiconductor classification. It plots the proportion of correctly identified metals (true positive rate) against the proportion of wrongly identified semiconductors (false positive rate) under different thresholds. }
  \label{fig:prediction}
\end{figure}

The database we used includes a diverse set of inorganic crystals ranging from simple metals to complex minerals. After removing ill-converged crystals, the full database has 46744 materials covering 87 elements, 7 lattice systems and 216 space groups. As shown in Figure \ref{fig:prediction}(a), the materials consist of as many as seven different elements, with 90\% of them binary, ternary and quaternary compounds. The number of atoms in the primitive cell ranges from 1 to 200, and 90\% of crystals have less than 60 atoms(Figure S2).  Considering most of the crystals originate from the Inorganic Crystal Structure Database (ICSD)\cite{hellenbrandt2004inorganic}, this database is a good representation of known stoichiometric inorganic crystals.

The CGCNN is a flexible framework that allows variance in the crystal graph representation, neural network architecture, and training process, resulting in different $f$ in Eq.~\ref{eq:optimization} and prediction performance. To choose the best model, we apply a train-validation scheme to optimize the prediction of formation energies of crystals. Each model is trained with 60\% of the data and then validated with 20\% of the data, and the best-performing model in validation set is selected. In our study, we find that the neural network architecture, especially the form of convolution function in Eq. \ref{eq:conv}, has the largest impact on prediction performance. We start with a simple convolution function,
\small
\begin{equation}\label{eq:conv_fun1}
  \bm{v}_i^{(t+1)} = g\left[ \left( \sum_{j, k} {\bm{v}_j^{(t)} \concat \bm{u}_{(i, j)_k}} \right) \bm{W}_c^{(t)} + \bm{v}_i^{(t)} \bm{W}_s^{(t)} + \bm{b}^{(t)} \right]
\end{equation}
\normalsize
where $\concat$ denotes concatenation of atom and bond feature vectors, $\bm{W}_c^{(t)}$, $\bm{W}_s^{(t)}$, $\bm{b}^{(t)}$ are the convolution weight matrix, self weight matrix, and bias of the $t$-th layer, respectively, and $g$ is the activation function for introducing non-linear coupling between layers. By optimizing hyperparameters in Table S1, the lowest mean absolute error (MAE) for the validation set is 0.108 eV/atom. One limitation of Eq. \ref{eq:conv_fun1} is that it uses a shared weight matrix $\bm{W}_c^{(t)}$ for all neighbors of $i$, which neglects the differences of interaction strength between neighbors. To overcome this problem, we design a new convolution function that first concatenates neighbor vectors $\bm{z}_{(i,j)_k}^{(t)} = \bm{v}_i^{(t)} \concat \bm{v}_j^{(t)} \concat \bm{u}_{(i, j)_k}$, then perform convolution by,
\begin{equation} \label{eq:conv_fun2}
\small
	\bm{v}_i^{(t+1)} = \bm{v}_i^{(t)} + \sum_{j, k} \sigma(\bm{z}_{(i,j)_k}^{(t)} \bm{W}_f^{(t)} + \bm{b}_f^{(t)}) \odot g(\bm{z}_{(i,j)_k}^{(t)} \bm{W}_s^{(t)} + \bm{b}_s^{(t)}) 
\end{equation}
where $\odot$ denotes element-wise multiplication and $\sigma$ denotes a sigmoid function. In Eq. \ref{eq:conv_fun2}, the $\sigma(\bm{\cdot})$ functions as a learned weight matrix to differentiate interactions between neighbors, and adding $\bm{v}_i^{(t)}$ makes learning deeper networks easier\cite{he2016deep}. We achieve MAE on the validation set of 0.039 eV/atom using the modified convolution function, a significant improvement compared to Eq. \ref{eq:conv_fun1}. In Figure S3, we compare the effects of several other hyperparameters on the MAE which are much smaller than the effect of convolution function.

Figure \ref{fig:prediction}(b)(c) shows the performance of the two models on 9350 test crystals for predicting the formation energy per atom. We find a systematic decrease of the mean absolute error (MAE) of the predicted values compared with DFT calculated values for both convolution functions as the number of training data is increased. The best MAE's we achieved with Eq. \ref{eq:conv_fun1} and Eq. \ref{eq:conv_fun2} are 0.136 eV/atom and 0.039 eV/atom, and 90\% of the crystals are predicted within 0.3 eV/atom and 0.08 eV/atom errors, respectively. In comparison, Kirklin \textit{et al.} reports that the MAE of DFT calculation with respect to experimental measurements in the Open Quantum Materials Database (OQMD) is 0.081--0.136 eV/atom depending on whether the energies of the elemental reference states are fitted, although they also find a large MAE of 0.082 eV/atom between different sources of experimental data. Given the comparison, our CGCNN approach provides a reliable estimation of DFT calculations and can potentially be applied to predict properties calculated by more accurate methods like \textit{GW}\cite{hybertsen1986electron} and quantum Monte Carlo\cite{foulkes2001quantum}.

\begin{table}
\caption{Summary of the prediction performance of seven different properties on test sets.\label{tab:predictions}}
\begin{ruledtabular}
\begin{tabular}{p{2cm}cccp{1.8cm}}
  Property    & \# of train data		&	Unit   & $\textrm{MAE}_{\textrm{model}}$   & $\textrm{MAE}_{\textrm{DFT}}$\\
  \hline
  Formation energy & 28046    		&  eV/atom  & 0.039 					&	0.081--0.136\cite{kirklin2015open} \\
  Absolute energy	& 28046			&  eV/atom	& 0.072					&	--\\
  Band gap			& 16458			&  eV		& 0.388					&	0.6\cite{jain2011high}\\
  Fermi energy		& 28046			&  eV		& 0.363					&	--\\
  Bulk moduli		& 2041 			&  log(GPa)	& 0.054					&	0.050\cite{de2015charting}\\
  Shear moduli		& 2041			& log(GPa)	& 0.087					&	0.069\cite{de2015charting}\\
  Poisson ratio		& 2041			& --			& 0.030					&	--\\

\end{tabular}
\end{ruledtabular}
\end{table}

After establishing the generality of CGCNN with respect to the diversity of crystals, we next explore its prediction performance for different material properties. We apply the same framework to predict the absolute energy, band gap, Fermi energy, bulk moduli, shear moduli, and Poisson ratio of crystals using DFT calculated data from the Materials Project\cite{jain2013commentary}. The prediction performance of Eq. \ref{eq:conv_fun2} is improved compared to Eq. \ref{eq:conv_fun1} for all six properties (Table S4). We summarize the performance in Table \ref{tab:predictions} and the corresponding 2D histograms in Figure S4. As we can see, the MAE of our model are close to or higher than DFT accuracy relative to experiments for most properties when ${\sim}10^4$ training data is used. For elastic properties, the errors are higher since less data is available, and the accuracy of DFT relative to experiments can be expected if ${\sim}10^4$ training data is available (Figure S5). 

Recently, Jong \textit{et al.}\cite{de2016statistical} developed a statistical learning (SL) framework using multivariate local regression on crystal descriptors to predict elastic properties using the same data from the Materials Project. By using the same number of training data, our model achieves root mean squared error (RMSE) on test sets of 0.105 log(GPa) and 0.127 log(GPa) for the bulk and shear moduli, which is similar to the RMSE of SL on the entire dataset of 0.0750 log(GPa) and 0.1378 log(GPa). Comparing the two methods, CGCNN predicts properties by extracting features only from crystal structure, while SL depends on crystal descriptors like cohesive energy and volume per atom. Recently, 1585 new crystals with elastic properties have been uploaded to the Materials Project database. Our model in Table \ref{tab:predictions} achieves MAE of 0.077 Log(GPa) for bulk moduli and 0.114 Log(GPa) for shear moduli on these crystals, showing good generalization to materials from potentially different crystal groups.

In addition to predicting continuous properties, CGCNN can also predict discrete properties by changing the output layer. By using a softmax activation function for the output layer and a cross entropy cost function, we can predict the classification of metal and semiconductor with the same framework. In Figure \ref{fig:prediction}(d), we show the receiver operating characteristic (ROC) curve of the prediction on 9350 test crystals. Excellent prediction performance is achieved with the area under the curve (AUC) at 0.95. By choosing a threshold of 0.5, we get metal prediction accuracy at 0.80, semiconductor prediction accuracy at 0.95, and overall prediction accuracy at 0.90.

\begin{figure}[tb]
  \centering
  \includegraphics[width=\linewidth]{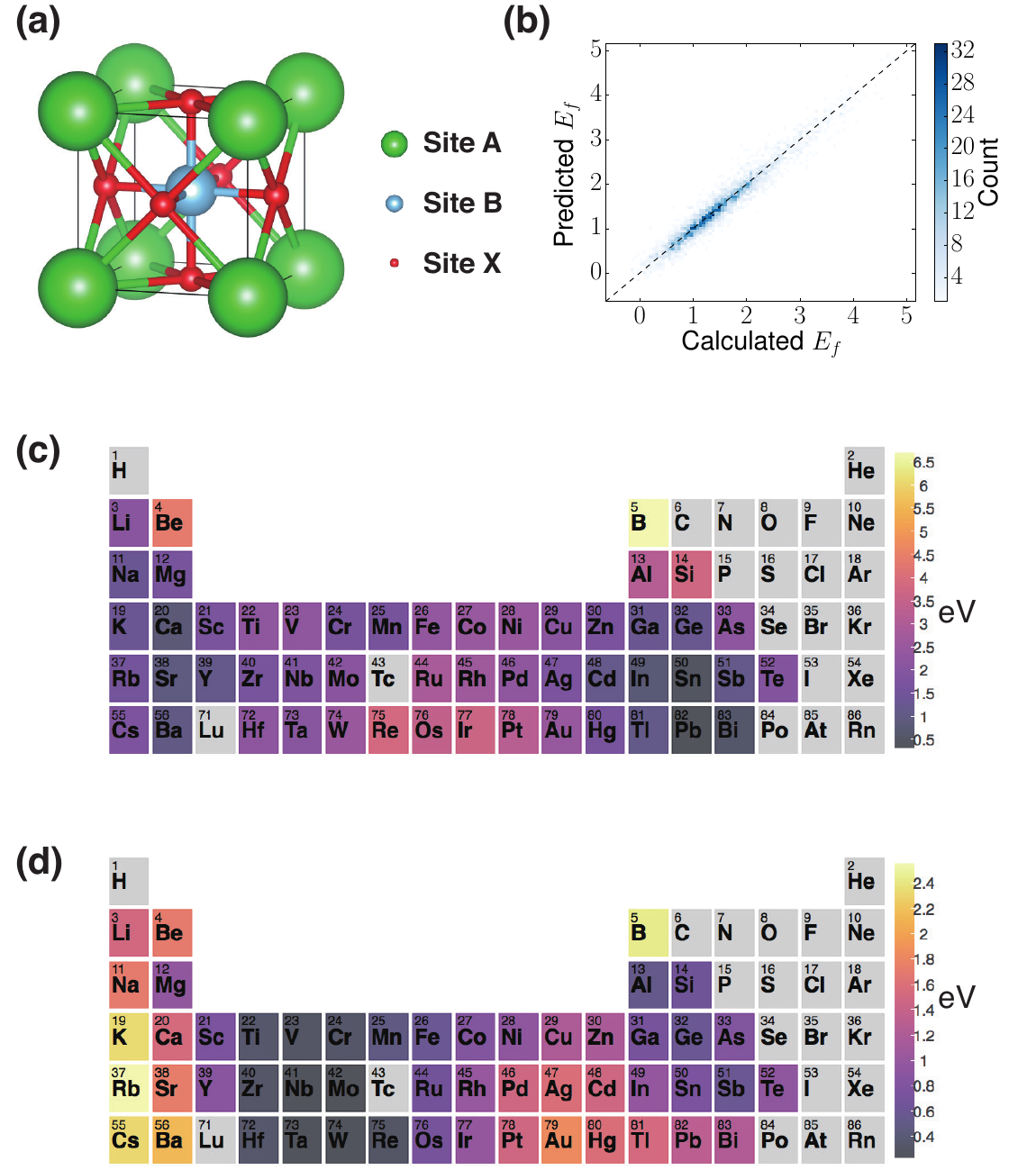}
  \caption{Extraction of site energy of perovskites from total energy above hull. (a) Structure of perovskites. (b) 2D histogram representing the predicted total formation against DFT calculated value. (c, d) Periodic table with the color of each element representing the mean of the site energy when the element occupies A site (c) or B site (d). }
  \label{fig:perovskite}
\end{figure}

Model interpretability is a desired property for any ML algorithms applied in materials science, because it can provide additional information for material design which may be more valuable than simply screening a large number of materials. However, non-linear functions are needed to learn the complex structure-property relations, resulting in ML models that are hard to interpret. CGCNN resolves this dilemma by separating the convolution and pooling layers. After the $R$ convolutional and $L_1$ hidden layers, we map the last atom feature vector $\bm{v}_i^{(R)}$ to a scalar $\tilde{v}_i$ and perform a linear pooling to predict the target property directly without the $L_2$ hidden layers (details discussed in SM). Therefore, we can learn the contribution of different local chemical environments, represented by $\tilde{v}_i$ for each atom, to the target property while maintaining a model with high capacity to ensure the prediction performance.

We demonstrate how these local chemical environment related information can be used to provide chemical insights and guide the material design by a specific example: learning the energy of each site in perovskites from the total energy above hull data. Perovskite is a crystal structure type with the form of \ce{ABX_3}, where the site A atom sits at a corner position, the site B atom sits at a body centered position and site X atoms sit at face centered positions (Figure \ref{fig:perovskite}(a)). The database\cite{castelli2012computational} we use includes the energy above hull of 18928 perovskite crystals, in which A and B sites can be any non-radioactive metals and X sites can be one or several elements from O, N, S, and F. We use the CGCNN with a linear pooling to predict the total energy above hull of perovskites in the database, using Eq. \ref{eq:conv_fun1} as the convolution function. The resulting MAE on 3787 test perovskites is 0.130 eV/atom as shown in Figure \ref{fig:perovskite}(b), which is slightly higher than using a complete pooling layer and $L_2$ hidden layers (0.099 eV/atom as shown in Figure S6) due to the additional constraints introduced by the simplified pooling layer. However, this CGCNN allows us to learn the energy of each site in the crystal while training with the total energy above hull, providing additional insights for material design.  

Figure \ref{fig:perovskite}(c, d) visualizes the mean of the predicted site energies when each element occupies the A and B site respectively. The most stable elements that occupy the A site are those with large radii due to the space needed for 12 coordinations. In contrast, elements with small radii like Be, B, Si are the most unstable for occupying the A site. For the B site, elements in groups 4, 5, and 6 are the most stable throughout the periodic table. This can be explained by crystal field theory, since the configuration of d electrons of these elements favors the octahedral coordination in the B site. Interestingly, the visualization shows that large atoms from groups 13-15 are stable in the A site, in addition to the well-known region of groups 1-3 elements. Inspired by this result, we applied a combinational search for stable perovskites using elements from group 13-15 as the A site and group 4-6 as the B site. Due to the theoretical inaccuracies of DFT calculations and the possibility of metastable phases that can be stabilized by temperature, defects, and substrates, many synthesizable inorganic crystals have positive calculated energies above hull at $\SI{0}{K}$. Some metastable nitrides can even have energies up to 0.2 eV/atom above hull as a result of the strong bonding interactions\cite{sun2016thermodynamic}. In this work, since some of the perovskites are also nitrides, we choose to set the cutoff energy for potential synthesizability at $\SI{0.2}{eV/atom}$. We discovered 33 perovskites that fall within this threshold out of 378 in the entire dataset, among which 8 are within the cutoff out of 58 in the test set (Table S5). Many of these compounds like \ce{PbTiO_3}\cite{shirane1952phase}, \ce{PbZrO_3}\cite{shirane1952phase}, \ce{SnTaO_3}\cite{lang2016improved}, and \ce{PbMoO3}\cite{takatsu2017cubic} have been experimentally synthesized. Note that \ce{PbMoO3} has calculated energy 0.18 eV/atom above hull, indicating that our choice of cutoff energy is reasonable. In general, chemical insights gained from CGCNN can significantly reduce the search space for high throughput screening. In comparison, there are only 228 potentially synthesizable perovskites out of 18928 in our database: the chemical insight increased the search efficiency by a factor of 7.

In summary, the crystal graph convolutional neural networks (CGCNN) presents a flexible machine learning framework for material property prediction and design knowledge extraction. The framework provides a reliable estimation of DFT calculations using around $10^4$ training data for 8 properties of inorganic crystals with diverse structure types and compositions. As an example of knowledge extraction, we apply this approach to the design of new perovskite materials and show that information extracted from the model is consistent with common chemical insights and significantly reduces the search space for high throughput screening. 

The code for the CGCNN is available from Ref.~\footnote{CGCNN website: \href{https://github.com/txie-93/cgcnn}{https://github.com/txie-93/cgcnn}}.

\begin{acknowledgements}
This work was supported by Toyota Research Institute. Computational support was provided through the National Energy Research Scientific Computing Center, a DOE Office of Science User Facility supported by the Office of Science of the U.S. Department of Energy under Contract No. DE-AC02-05CH11231, and the Extreme Science and Engineering Discovery Environment, supported by National Science Foundation grant number ACI-1053575.
\end{acknowledgements}

\bibliography{graph_fp.bib}

\widetext
\clearpage

\begin{center}
\textbf{\large Supplemental Materials: Crystal Graph Convolutional Neural Networks for an Accurate and Interpretable Prediction of Material Properties}
\end{center}
\setcounter{equation}{0}
\setcounter{figure}{0}
\setcounter{table}{0}
\setcounter{page}{1}
\makeatletter
\renewcommand{\theequation}{S\arabic{equation}}
\renewcommand{\thefigure}{S\arabic{figure}}
\renewcommand{\bibnumfmt}[1]{[S#1]}
\renewcommand{\citenumfont}[1]{S#1}

\section{Supplementary Methods and Discussions}

\subsection{Construction of crystal graphs}

The connectivity between atoms in a crystal graph is determined by a method inspired by Ref.~\cite{isayev2017universal}. For each atom, neighbors are first searched within $\SI{6}{\angstrom}$ radius, and they are considered as connected when they share a Voronoi face~\cite{blatov2004voronoi} with the center atom and have interatomic distance lower than the sum of the Cordero covalent bond lengths~\cite{cordero2008covalent} with a $\SI{0.25}{\angstrom}$ tolerance. Therefore, only strong bonding interactions are considered in the crystal graph construction.

It is worth noting that by using convolution function Eq.~5 in the main text, the connectivity between atoms becomes less important because the $\sigma(\bm{\cdot})$ part automatically ignores weak bonds. In practice, we discover that connecting 12 nearest neighbors in the initial graph construction performs as good as using the method described above.

 Atom and bond properties are encoded in node feature vectors $\bm{v}_i$ and edge feature vectors $\bm{u}_{(i, j)_k}$ using one hot encoding. For discrete values, the vectors are encoded according to the category that the value belongs to; for continuous values, the range of property values is evenly divided to 10 categories and the vectors are encoded accordingly. The full list of atom and bond properties as well as their ranges are in Table \ref{tab:atom-feature} and Table \ref{tab:bond-feature}. For instance, if we use the group number and period number as atom features, the atom feature vector for H will be a 27-dimensional vector with 1st and 19th element being 1 and other elements being 0. If the interatomic distance is 0.7, then the bond feature vector will be a 10-dimensional vector with 1st element being 1 and other elements being 0.

\subsection{Illustrative example for differentiating NaCl and KCl}

We provide a simple illustrative example to explain how CGCNN works by completing the task of differentiating the structures of NaCl and KCl. Concretely, the task is to predict +1 if the crystal structure is NaCl and -1 if the structure is KCl. We can accomplish it with a CGCNN with only one convolutional layer and one pooling layer.

\begin{figure}[htb]
    \centering
    \includegraphics[width = 0.9 \textwidth]{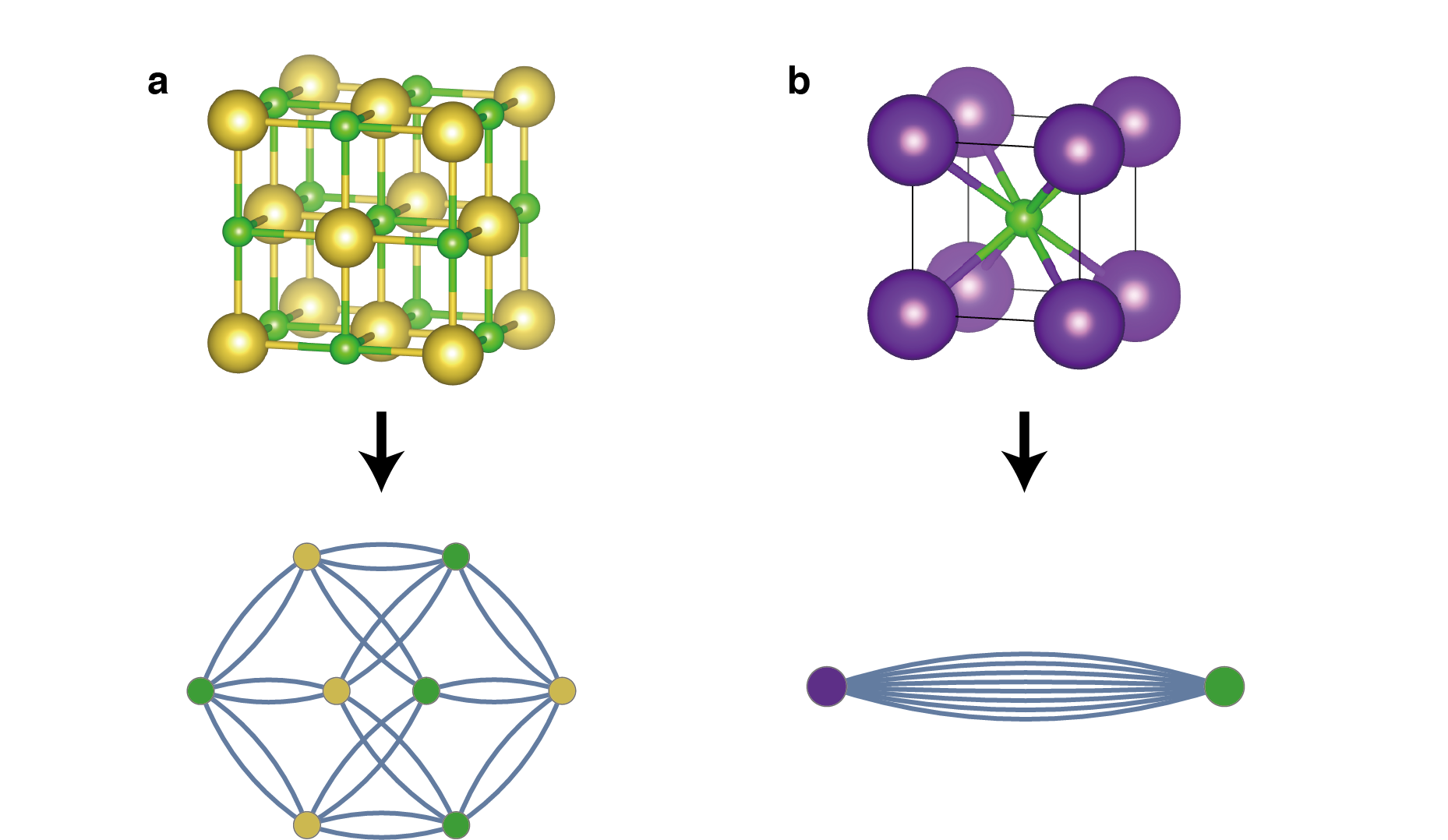}
    \caption{The crystal structures and crystal graphs of NaCl (a) and KCl (b). }
    \label{fig:nacl-kcl}
\end{figure}

In Fig.~\ref{fig:nacl-kcl}, we convert the original crystal structures of NaCl and KCl to their corresponding crystal graphs. Note that the four Na and Cl nodes in the crystal graph of NaCl all have the same chemical environments, and we can actually simplify it to a crystal graph with two nodes. Here we use this eight-node crystal graph to illustrate that CGCNN is invariant to the choice of unit cells. For simplicity, we only use atom feature vectors without bond feature vectors in crystal graphs, since it is enough to differentiate NaCl and KCl. Specifically, each node is represented by following vectors according to the element information.

\begin{equation}
  \bm{v}_{\mathrm{Cl}} = \begin{pmatrix} 1 & 0 & 0 \end{pmatrix};
  \bm{v}_{\mathrm{Na}} = \begin{pmatrix} 0 & 1 & 0 \end{pmatrix};
  \bm{v}_{\mathrm{K}} = \begin{pmatrix} 0 & 0 & 1 \end{pmatrix};
\end{equation}

Then, we apply one convolutions to each node in the two crystal graphs according to Eq.~4 in the main text. We initialize $\bm{W}_c$ and $\bm{W}_s$ as,
\begin{equation}
  \bm{W}_c = \begin{pmatrix} w_{c1} \\ w_{c2} \\ w_{c3} \end{pmatrix};
  \bm{W}_s = \begin{pmatrix} w_{s1} \\ w_{s2} \\ w_{s3} \end{pmatrix};
\end{equation}
and for simplicity, we set $\bm{b} = \bm{0}$ and $g(x) = x$. After one convolution, in NaCl, the feature of each node becomes,
\begin{align}
  \bm{v}_{\mathrm{Na}}^{(1)} &= 6 w_{c1} + w_{s2} \\
  \bm{v}_{\mathrm{Cl}}^{(1)} &= 6 w_{c2} + w_{s1}
\end{align}
and in KCl,
\begin{align}
  \bm{v}_{\mathrm{K}}^{(1)} &= 8 w_{c1} + w_{s3} \\
  \bm{v}_{\mathrm{Cl}}^{(1)} &= 8 w_{c3} + w_{s1}
\end{align}

After convolution, we apply a simple normalized pooling by summing over the feature vectors of all nodes $\bm{v}_{i}^{(1)}$ within the crystal graph, and then divide it by the total number of nodes. Consequently, the overall feature vectors of the two crystal graphs are,
\begin{align}
  \bm{v}_{\mathrm{NaCl}} = 3 w_{c1} + 3 w_{c2} + 0.5w_{s1} + 0.5w_{s2} \\
  \bm{v}_{\mathrm{KCl}} =  4 w_{c1} + 4 w_{c3} + 0.5w_{s1} + 0.5w_{s3}
\end{align}

Since the crystal feature vectors are already one-dimension, we do not need another output layer to map them into the target value. So, our predictions are,
\begin{align}
  \hat{y}_{\mathrm{NaCl}} = 3 w_{c1} + 3 w_{c2} + 0.5w_{s1} + 0.5w_{s2}\label{eq:nacl} \\
  \hat{y}_{\mathrm{KCl}} =  4 w_{c1} + 4 w_{c3} + 0.5w_{s1} + 0.5w_{s3}\label{eq:kcl}
\end{align}

We can easily find $\bm{W}_c$ and $\bm{W}_s$ that make $\hat{y}_{\mathrm{NaCl}} = 1$ and $\hat{y}_{\mathrm{KCl}} = -1$, showing that CGCNN are capable of differentiating NaCl and KCl. In this example, the existence of certain weights $w_{ci}$ or $w_{si}$ in Eq.~\ref{eq:nacl} and Eq.~\ref{eq:kcl} indicates the occurrence of elements as centers or neighbors respectively, while the factor represents the frequency of occurrence, both of which can help differentiating the two crystals.

When more training data is available, we could not find $\bm{W}_c$ and $\bm{W}_s$ that result in zero loss since the weights are shared for all crystals. Methods like stochastic gradient descent (SGD) can me used to minimize the loss and find the weights that maximize the prediction performance. Also, more complex CGCNN structures can be used to capture more structure information.

\subsection{Hyperparameter optimization}\label{sec:hyper}

The hyperparameters are parameters that defines the CGCNN. They include graph parameters that are used to generate the crystal graph, architecture parameters that are used to define the convolutional neural network on top of the crystal graph, and training parameters that are used for the training process. Unlike the weights that are trained via SGD, the hyperparameters are chosen through a train-validation process.

We first randomly divide our database into three parts: training set (60\%), validation set (20\%), and test set (20\%). Model with different hyperparameters are trained on training set via SGD, and the resulting weights are used to predict the property of crystals in validation set. By comparing to the properties calculated by DFT, and the hyperparameters that provide lowest MAE or AOC in the validation set are chosen as the optimum hyperparameters.

\begin{figure}[htb]
    \centering
    \includegraphics[width = \textwidth]{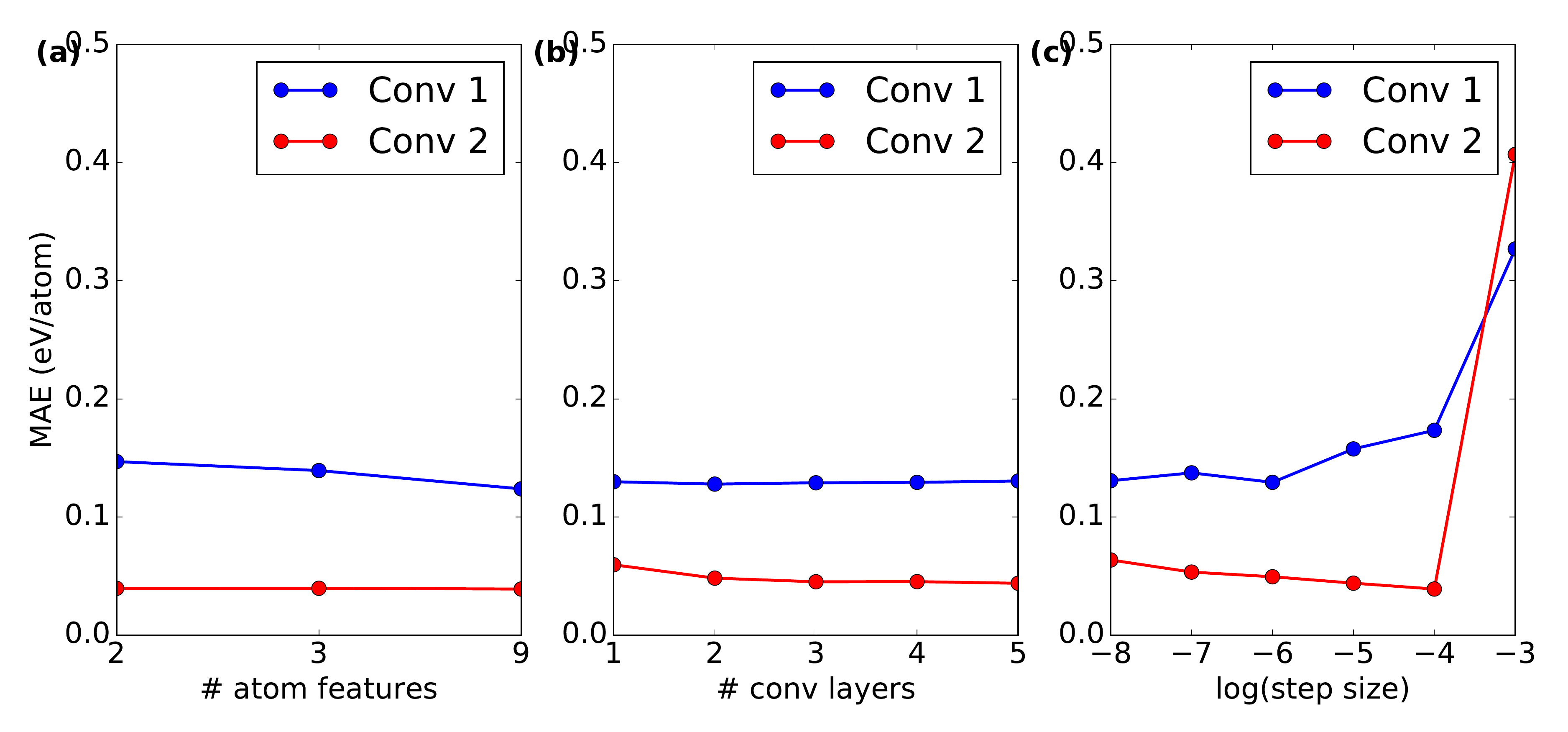}
    \caption{The effect of different hyperparameters on the validation mean absolute errors (MAEs). The blue points denotes models using convolution function Eq.~4, and the red points denotes models using Eq.~5. (a) Number of atom features. The 2 features include group number and period number, the 3 features additionally include electronegativity, and the 9 features include all properties in Table \ref{tab:atom-feature}. (b) Number of convolutional layers. (c) Logarithm of the step size. }
    \label{fig:hyper}
\end{figure}

For convolution function Eq.~4, we use random search to optimize all the hyperparameters listed in Table \ref{tab:list-of-hyper}. Since the influence of hyperparameter optimization is much smaller than the convolution functions as shown in Fig.~\ref{fig:hyper}, we only optimize number of convolutional layers, regularization term, and step size of the Adam optimizer when using convolution function Eq.~5.

\subsection{Pooling layer choices}
Throughout this work, normalized summation is used as a pooling function as described in the main text. However, for convolutional function Eq.~4, we sum up different feature vectors for the purpose of either maximizing prediction performance or interpretability.

To maximize prediction performance, we utilize feature vectors from all convolutional layers. Each feature vector $\bm{v}_i^{(t)}$ from the $t$-th convolution is first transformed to the same dimension by a linear map and then sparsified by a Softmax function (Eq.~\ref{eq:softmax}), resulting in a vector $\bm{\tilde{v}}_i^{(t)}$ with the same dimension. Then, the pooling layer sums over feature vectors $\bm{\tilde{v}}_i^{(t)}$ from all convolutional layers where $t=0, 1, .., R$ and over all atoms $i = 1, 2, ..., N$. The resulting crystal vector $\bm{v_c}$ is then normalized to be invariant to crystal size. This pooling includes all intermediate feature vectors representing chemical environments of various radius, which is more complete and resulting better prediction performance.


\begin{equation}
  \textrm{Softmax} \left(\bm{z} \right)_j = \frac{\exp(z_j)}{\sum_k \exp(z_k)}
\end{equation}

\begin{equation}\label{eq:softmax}
    \bm{\tilde{v}}_i^{(t)} = \textrm{Softmax} ( \bm{W}_{t}\bm{v}_i^{(t)} + \bm{b}_{t} )
\end{equation}

\begin{equation}
  \bm{v_c} = \sum_{t, i} \bm{\tilde{v}}_i^{(t)}
\end{equation}

To get better interpretability, we directly map the final feature vector after $R$ convolutional layers and $L_1$ hidden layers $\bm{v}_i^{(R)}$ to a scalar $\tilde{v}_i$ using a linear transform (Eq.~\ref{eq:linear}), and then summing up the scalars to predict the target property $v_c$. After training with the global property $v_c$, the model automatically learns the contribution of each local chemical environments represented by $\tilde{v}_i$.

\begin{equation}\label{eq:linear}
    \tilde{v}_i =  \bm{W}^T \bm{v}_i^{(R)} + b
\end{equation}

\begin{equation}
  v_c = \frac{1}{N} \sum_{i} \tilde{v}_i
\end{equation}

For convolution function Eq.~5, we directly compute the normalized sum of the last feature vector $\bm{v}_i^{(R)}$ after R convolutional layers as a pooling function (Eq.~\ref{eq:mean}) since the residual structure in Eq.~5 allows information passed to the last layer which makes adding $\bm{v}_i^{(t)}$ from previous convolutional layers unnecessary. 

\begin{equation}\label{eq:mean}
	\bm{v}_c = \frac{1}{N} \sum_{i} \bm{v}_i^{(R)}
\end{equation}

\clearpage
\section{Supplementary Figures}

\begin{figure}[htb]
    \centering
    \includegraphics[width = 0.6 \textwidth]{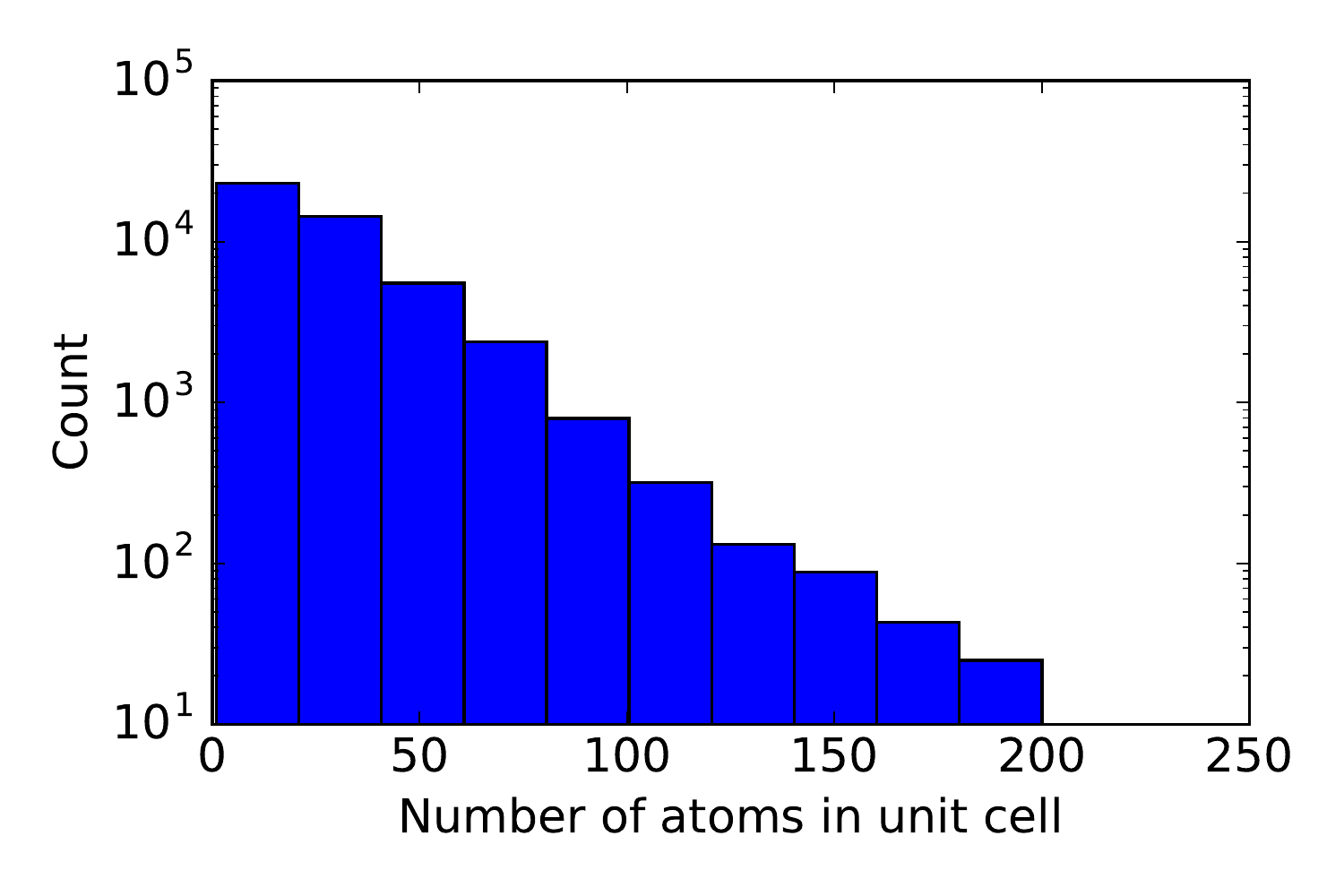}
    \caption{Histogram representing the distribution of the number of atoms in the primitive cells. }
    \label{fig:matproj}
\end{figure}

\begin{figure}[htb]
    \centering
    \includegraphics[width = \textwidth]{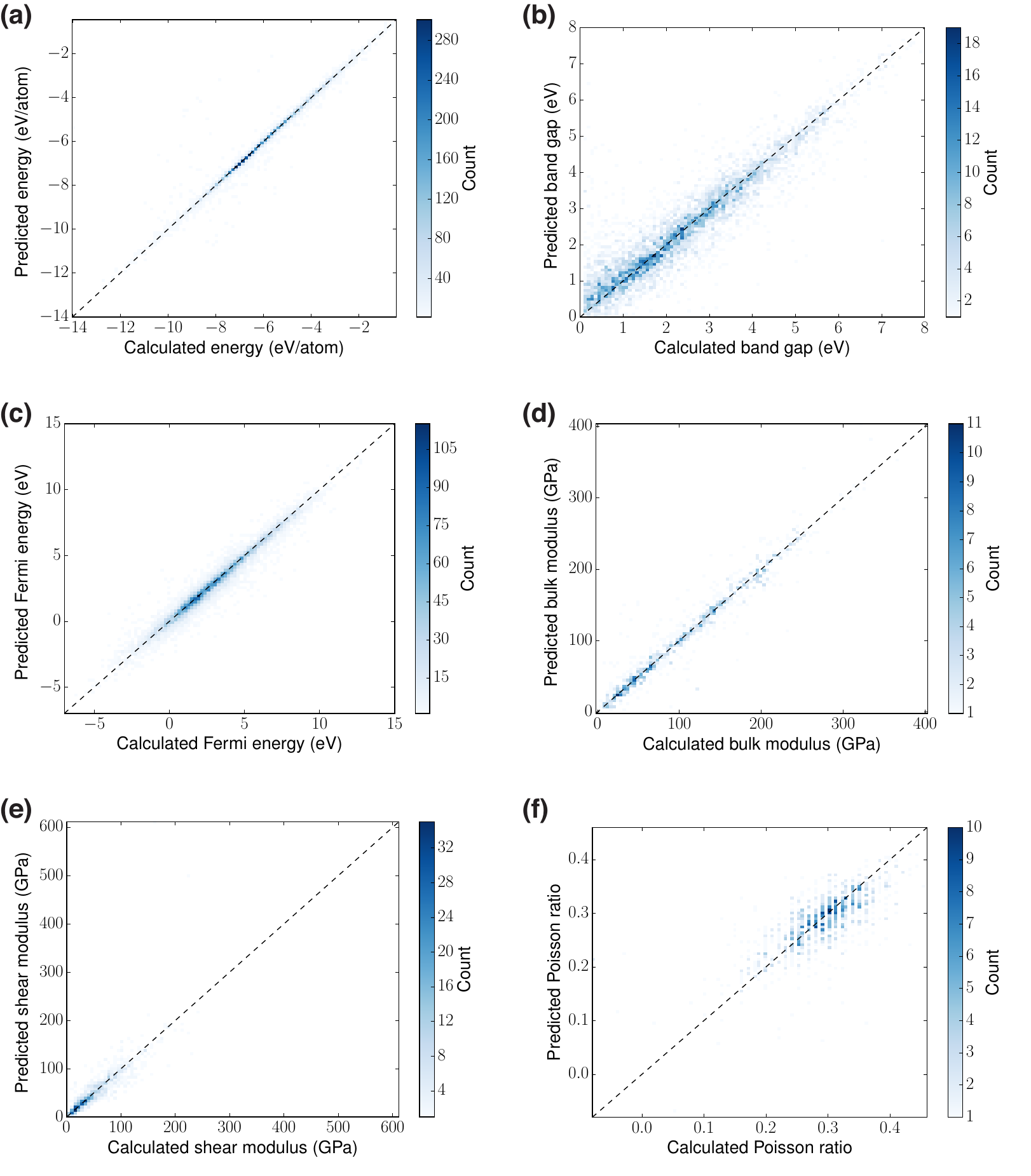}
    \caption{2D histogram visualizing the predictive performance of six properties. (a) Total energy per atom. (b) Band gap. (c) Fermi energy. (d) Bulk moduli. (e) Shear moduli. (e) Poisson ratio. }
    \label{fig:other-properties}
\end{figure}

\begin{figure}[htb]
    \centering
    \includegraphics[width = 0.6 \textwidth]{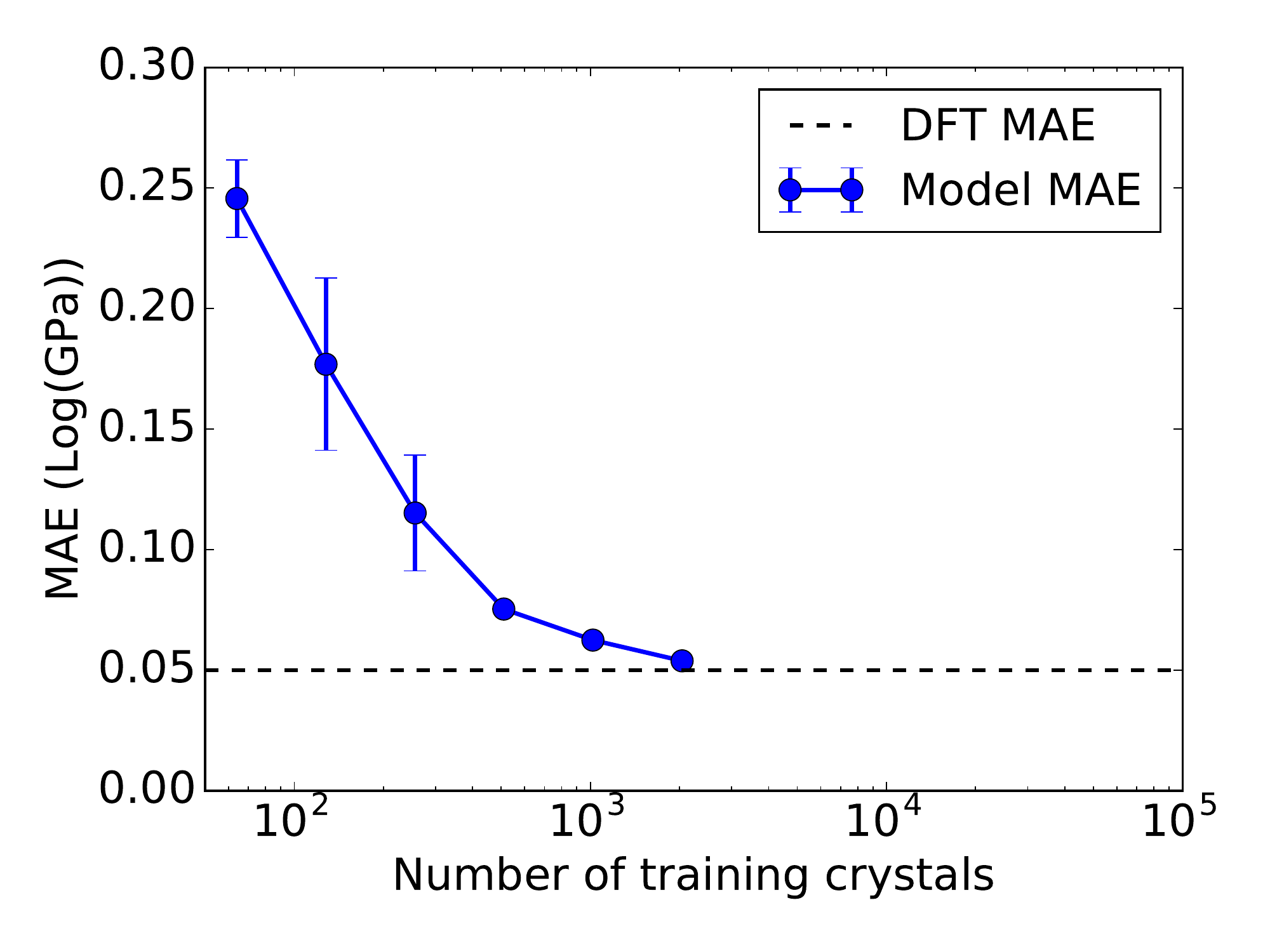}
    \caption{The MAE of predicted bulk modulus with respect to DFT values against the number of training crystals. The dashed line shows the MAE of DFT calculations with respect to experimental results~\cite{de2015charting}, which is 0.050 Log(GPa). }
    \label{fig:bulk-size-trend}
\end{figure}

\begin{figure}[htb]
    \centering
    \includegraphics[width = 0.6 \textwidth]{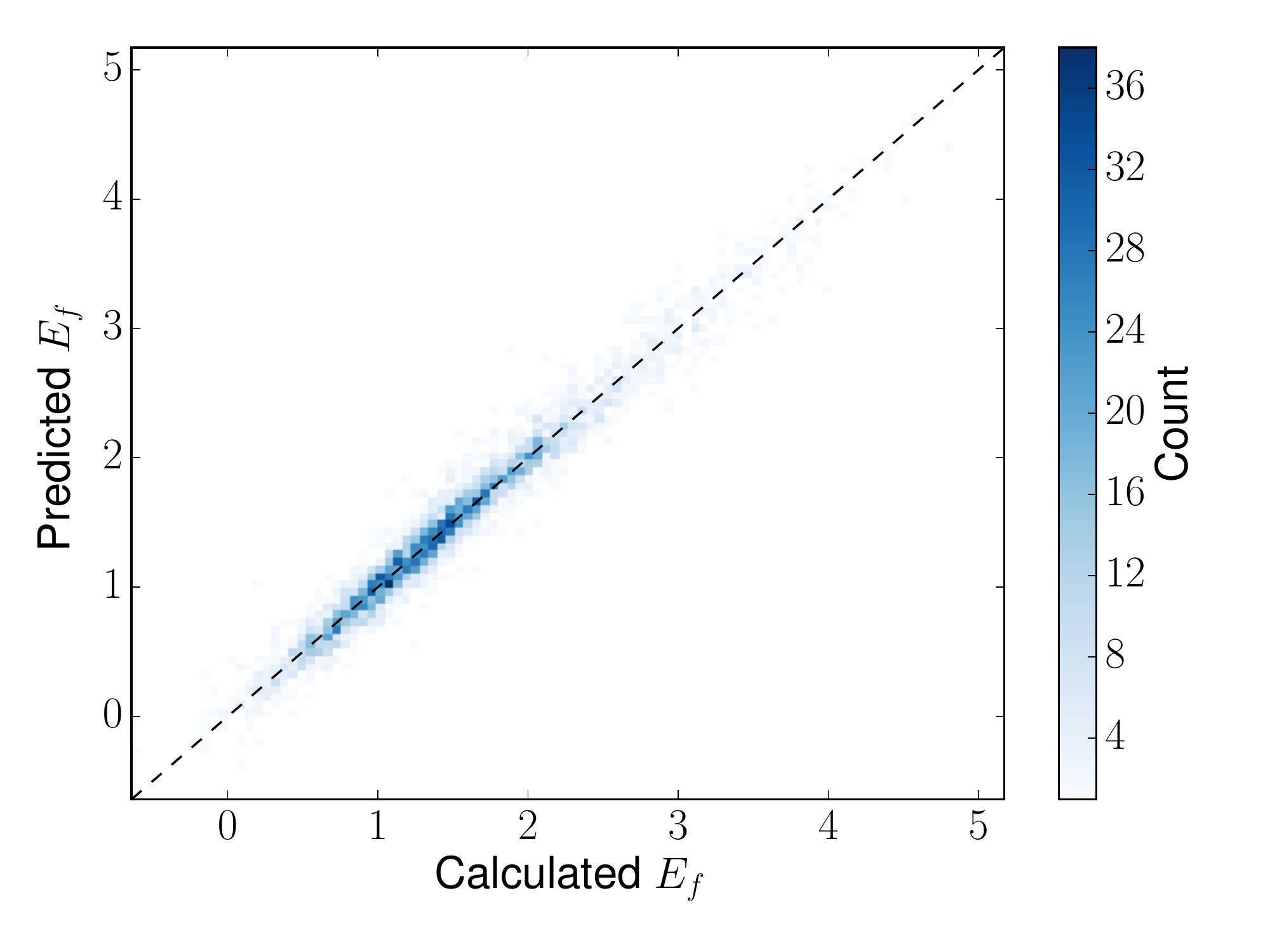}
    \caption{2D histogram visualizing the performance of predicting formation energy of perovskites using a full pooling layer with Eq.~4 as the convolution function.  }
    \label{fig:perovskite-full}
\end{figure}

\clearpage

\section{Supplementary Tables}

\begin{table}[htb]
\caption{A list of hyperparameters that are optimized in this work. }
\begin{ruledtabular}
\begin{tabular}{cc}
  Hyperparameter                                  & Range		\\
  \hline
  Number of properties used in atom feature vector $\bm{v}_i$ & 2, 3, 10\\
  Number of convolutional layers                  & 1, 2, ..., 5    	\\
  Length of learned atom feature vector $\bm{v}_i^{(t)}$  & 10, 20, 50, 100, 200 \\
  Number of hidden layer $L_1$\footnotemark[1]    & 1, 2, ..., 4 \\
  Number of hidden layer $L_2$\footnotemark[1]    & 1, 2, ..., 5 \\
  Regularization term $\lambda$\footnotemark[2]   & $\me^{-6}, \me^{-4}$, $\me^{-2}$, $\me^{0}$ \\
  Scaling factor of the Gaussian initialization of weights  & $\me^{-8}$, $\me^{-6}$, $\me^{-4}$, $\me^{-2}$ \\
  Step size of the Adam optimizer\cite{kingma2014adam}  &  $\me^{-8}$, $\me^{-7}$, $\me^{-6}$, $\me^{-5}$, $\me^{-4}$, $\me^{-3}$ \\
  Dropout fraction\cite{srivastava2014dropout}   & 0, 0.1, 0.2 \\

\end{tabular}
\end{ruledtabular}
\footnotetext[1]{The hidden layers $L_1$ and $L_2$ are not used simultaneously in this work. We only use $L_2$ for the prediction of material properties and $L_1$ for the learning of individual site energies in perovskites. }
\footnotetext[2]{$L^2$ regularization term $\lambda \left\Vert \bm{W} \right\Vert^2_2$ is added to cost function to reduce overfitting. }
\label{tab:list-of-hyper}
\end{table}

\begin{table}[htb]
\caption{Properties used in atom feature vector $\bm{v}_i$}
\begin{ruledtabular}
\begin{tabular}{cccc}
  Property        & Unit		&	Range           & \# of categories    \\
  \hline
  Group number    & --    	&  1,2, ..., 18   & 18 					\\
  Period number	& --			&  1,2, ..., 9\footnotemark[1]	& 9					\\
  Electronegativity\cite{sanderson1951interpretation, sanderson1952explanation}	& --			&  0.5--4.0		& 10					\\
  Covalent radius\cite{cordero2008covalent}		& pm		&  25--250		& 10					\\
  Valence electrons		& -- 			&	1, 2, ..., 12	& 12					\\
  First ionization energy\cite{kramida2013nist}\footnotemark[2]		& eV			& 1.3--3.3		& 10					\\
  Electron affinity\cite{haynes2014crc}		& eV			& -3--3.7			& 10					\\
  Block     & --            & s, p, d, f    &   4   \\
  Atomic volume\footnotemark[2] & $\SI{}{cm^3/mol}$ & 1.5--4.3  & 10  \\

\end{tabular}
\end{ruledtabular}
\footnotetext[1]{The lanthanide and actinide elements are considered as period 8 and 9 respectively.}
\footnotetext[2]{Log scale is used for these properties. }
\label{tab:atom-feature}
\end{table}

\begin{table}[htb]
\caption{Properties used in bond feature vector $\bm{u}_{(i, j)_k}$}
\begin{ruledtabular}
\begin{tabular}{cccc}
  Property        & Unit		&	Range           & \# of categories    \\
  \hline
  Atom distance    & $\SI{}{\angstrom}$    	&  0.7--5.2   & 10 					\\

\end{tabular}
\end{ruledtabular}
\label{tab:bond-feature}
\end{table}

\begin{table}[htb]
\caption{Comparison of the prediction performance of seven different properties on test sets using different convolution functions.}
\begin{ruledtabular}
\begin{tabular}{ccccc}
  Property    & \# Train data		&	Unit   & $\textrm{MAE}_{\textrm{Eq.~4}}$   & $\textrm{MAE}_{\textrm{Eq.~5}}$ \\
  \hline
  Formation energy & 28046    		&  eV/atom  & 0.112 					&	0.039\\
  Absolute energy	& 28046			&  eV/atom	& 0.227					&	0.072\\
  Band gap			& 16458			&  eV		& 0.530					&	0.388\\
  Fermi energy		& 28046			&  eV		& 0.564					&	0.363\\
  Bulk moduli		& 2041 			&  Log(GPa)	& 0.084					&	0.054\\
  Shear moduli		& 2041			&  Log(GPa) & 0.113					&	0.087\\
  Poisson ratio		& 2041			& --			& 0.033					&	0.030\\

\end{tabular}
\end{ruledtabular}
\label{tab:general-param}
\end{table}

\squeezetable
\squeezetable

\begin{table}[h]
\caption{Perovskites with energy above hull lower than 0.2 eV/atom discovered using combinational search. }
\begin{ruledtabular}
\begin{tabular}{cccc}
  Formula &   A site   & B site   & Formation energy per atom (eV/atom)   \\
  \hline

\multicolumn{4}{c}{Training Set (60\%)} \\
  \hline
  TlNbO3 & Tl & Nb & 0.0 \\
  SnTiO3 & Sn & Ti & 0.1 \\
  PbVO3 & Pb & V & 0.04 \\
  SnTaO3 & Sn & Ta & 0.0 \\
  TlWO3 & Tl & W & 0.12 \\
  PbMoO3 & Pb & Mo & 0.18 \\
  PbCrO3 & Pb & Cr & 0.14 \\
  SnNbO3 & Sn & Nb & 0.14 \\
  SnTaO2N & Sn & Ta & 0.14 \\
  TlTaOFN & Tl & Ta & 0.18 \\
  TlTaO2F & Tl & Ta & 0.04 \\
  TlHfO2F & Tl & Hf & 0.18 \\
  PbTiO3 & Pb & Ti & 0.06 \\
  InNbO3 & In & Nb & 0.06 \\
  InWO3 & In & W & 0.18 \\
  InTaO3 & In & Ta & -0.16 \\
  InNbO2F & In & Nb & 0.18 \\
  InTaO2S & In & Ta & 0.18 \\
\hline
\multicolumn{4}{c}{Validation Set (20\%)} \\
  \hline
  TlNbO2F & Tl & Nb & 0.08 \\
  TlZrO2F & Tl & Zr & 0.14 \\
  SnVO3 & Sn & V & 0.12 \\
  TlTiO2F & Tl & Ti & -0.02 \\
  PbNbO2N & Pb & Nb & 0.18 \\
  PbZrO3 & Pb & Zr & 0.08 \\
  PbNbO3 & Pb & Nb & 0.04 \\

\hline
\multicolumn{4}{c}{Test Set (20\%)} \\
  \hline
  BiCrO3 & Bi & Cr & 0.14 \\
  PbVO2F & Pb & V & 0.14 \\
  SnNbO2N & Sn & Nb & 0.18 \\
  PbHfO3 & Pb & Hf & 0.1 \\
  TlTaO3 & Tl & Ta & 0.1 \\
  PbTaO3 & Pb & Ta & 0.18 \\
  InTaO2F & In & Ta & 0.08 \\
  InZrO2F & In & Zr & 0.16 \\

\end{tabular}
\end{ruledtabular}
\label{tab:stable-combinational}
\end{table}

\end{document}